\documentclass[preprint,10pt]{elsarticle}

\usepackage{amssymb}
\usepackage{amsfonts}   
\usepackage{amsmath}
\usepackage{multirow} 
\usepackage[dvipsnames]{xcolor}
\usepackage[normalem]{ulem}
\usepackage{verbatim}
\usepackage{subcaption}
\usepackage{graphicx}
\usepackage{tikz}
\usetikzlibrary{decorations.pathreplacing,calligraphy}
\newtheorem{definition}{Definition}[section]

\newtheorem{example}{Example}[section]

\begin{document}


\begin{frontmatter}



\title{Sorting ECGs by lag irreversibility}

\author[ipicyt]{Nazul Merino Negrete}
\author[ipicyt]{Cesar Maldonado}
\address[ipicyt]{IPICyT/Divisi\'{o}n de Control y Sistemas Din\'{a}micos, Camino a la Presa San Jos\'{e} 2055. Lomas 4a. secci\'{o}n. San Luis Potos\'{i} S.L.P. M\'{e}xico}

\author[uaem]{Ra\'{u}l Salgado-Garc\'{i}a}
\address[uaem]{Centro de Investigaci\'{o}n en Ciencias-IICBA, Universidad Aut\'{o}noma del Estado de Morelos. Avenida Universidad 1001, colonia Chamilpa, 62209, Cuernavaca, M\'{e}xico.}



\begin{abstract}
In this work we introduce the lag irreversibility function as a method to assess time-irreversibility in discrete time series. It quantifies the degree of time-asymmetry for the joint probability function of the state variable under study and the state variable lagged in time. We test its performance in a time-irreversible Markov chain model for which theoretical results are known. Moreover, we use our approach to analyze electrocardiographic recordings of four groups of subjects: healthy young individuals, healthy elderly individuals, and persons with two different disease conditions, namely, congestive heart failure and atrial fibrillation. We find that by studying jointly the variability of the amplitudes of the different waves in the electrocardiographic signals, one can obtain an efficient method to discriminate between the groups already mentioned. Finally, we test the accuracy of our method using the ROC analysis.


\end{abstract}


\end{frontmatter}


\section{Introduction}
\label{sec:introduction}

In recent years, assessing the time-reversal asymmetry has become an im\-por\-tant tool for the analysis of time series ~\cite{Zanin2021Algorithmic}. This is because the irreversibility of a (stochastic or deterministic) process is a feature that provides information about the nature of the underlying dynamics~\cite{Porporato2007Irreversibility, gaspard2004time, latora1999kolmogorov}. For instance, the time-irreversibility is closely related to the entropy production in physical systems~\cite{Landi2021Irreversible}, a signature of out-of-equilibrium processes~\cite{Gaspard2022Book, Landi2021Irreversible, Roldan2012Entropy}. Also, the time-reversal asymmetry is a property related to the presence of non-linear correlations as well as the presence of non-Gaussian fluctuations, among other interesting proper\-ties~\cite{Daw2000Symbolic}. 

Several approaches has been proposed to determine the irreversibility of a time series~\cite{Zanin2021Algorithmic, kathpalia2021time, Zanin2021Assesing}. For example, a natural way to assess irreversibility is comparing directly the joint distribution of the process forward and backward in time~\cite{Porporato2007Irreversibility, diks1995reversibility}. This is usually done by encoding the time series into a symbolic sequence and then computing the occurrence of words and their respective time-reversed versions~\cite{Daw2000Symbolic}. Another well-established method to quantify the degree of irreversibility in symbolic sequences is to compute the Kullback-Leibler divergence between the process forward in time and its time-reversal version~\cite{ Roldan2012Entropy, Porporato2007Irrev, Roldan2010Estimating}. Many other approaches that is worth to mention are: visibility graph approach~\cite{lacasa2012time}, ordinal patterns analysis~\cite{martinez2018Detection}, recurrence-time statistics~\cite{ChR, Salgado2021Estimating, Yun2008Estimating} and  recurrence plots~\cite{guzik2006heart}, just to mention a few examples.

Since time irreversibility is a common feature of the out-of-equilibrium pheno\-mena, it has been widely studied in physics, biology, mathe\-matics, chemistry, economic sciences, among other disciplines~\cite{ Zanin2021Algorithmic, Landi2021Irreversible, Gaspard2022Book, irreversibility2016Flanagan,  jiang2004mathematical}. For instance, time-irreversibility has been implemented in the analysis of financial time series in~\cite{irreversibility2016Flanagan, Li2018TimeIrreversibility, xia2014classifying}. Using a permutation patterns approach,  in~\cite{Zanin2021Assesing}, M. Zanin stu\-died the irreversibility properties of different chaotic dynamical systems. In~\cite{Porporato2007Irrev}, Porporato \textit{et al.,} estimated the irreversibility by means of the Kullback-Leibler divergence of a stochastic process defined by a combination of an asym\-me\-tric jumps Poisson process and an Ornstein-Uhlenbeck process.  They later applied that same methodology to quantify the asymmetry of the time series of the discharge measurements for the Po River in Italy. With respect to the physics literature there is a huge amount of studies concerning the relation with the fluctuation-dissipation theorem, as well as for applications in out-of-equilibrium systems.
For a comprehensive study on irreversibility in statistical mechanics, see~\cite{Gaspard2022Book} and references therein. The irreversibility of biological and physiological signals has recently arisen interest. For instance, there are studies in electroencephalo\-gra\-phic time series~\cite{zanin2020time}, in Markovian models of Spike Trains~\cite{CofreM}, and also it has been shown that the irreversibility ana\-lysis can be useful for discriminating between coding and non-coding DNA sequences~\cite{Salgado2021Time, Salgado2021Estimating}.

Here we focus in the analysis of irreversibility of time series obtained from some properties (derived signals) of electrocardiograms (ECG's). 
In that context, for instance, the temporal asymmetry of heart rate variability signals has been studied in~\cite{Cammarota2007Time} by Cammarota \textit{et al.},  using a ternary symbolization of the RR-intervals, which was later extended in \cite{Hou2013Combination}. Using indices of irreversibility such as Poincar\'{e} plots and related methods, it has been found that the heart rate variability signals display a temporal asymmetry~\cite{Piskorski2007Geometry, Karmakar2009Defining,Yan2017Area}. In~\cite{Costa2005broken},  Costa \textit{et al.}, suggested that loss of irreversibility might be related to aging and disease of individuals. They analyzed the signal using different scales of the time series and estimated the irreversibility of the probability density function of the system and its time reversal, which considers the energy and heat fluxes in the process. Related methods, such as estimation of the change of entropy in (a so-called) natural time are given by~\cite{Varostos2007Identifying} and ~\cite{Sarlis2015Change}. Recently, in~\cite{Nazul}, under an assumption of Markovianity of the signals, the authors estimated the entropy production of the symbolized time series for both the electrocardiographic signals and the heart rate variability of three groups of patients; healthy ones, patients with Atrial Fibrillation and individuals with Congestive Heart Failure. There, it is shown that it is possible to discriminate the healthy ones from the diseased groups by means of the entropy production. Other studies that analyze time-irreversibility  in order to determine the health condition of individuals from the electrocardiographic time series can be found in~\cite{porta2008temporal, casali2008multiple, porta2009assessment, hou2010analysis, piskorski2011asymmetric,Karmakar2015}. 

In the present work we define an index to quantify the time irreversibility, which we call \textit{lag irreversibility function}.  
This quantity can be applied not only to single-signal processes, but to processes possessing more than one signal, i.e., the lag irreversibility function can be evaluated for multivariate time series. 
This is important in applications, for instance, we apply our irreversibility index to ECG signals, since it takes into account information on different stages (or features) of the ECG through their corresponding heart rate variability signals. This may contribute to get a better estimate of the time-reversal asymmetry using  not so long recordings. So, by computing the lag irreversibility function we are able to discriminate signals for four groups of individuals with different health conditions, including healthy elderly patients. We test the accuracy of our method using the ROC analysis.

The article is organized as follows. In Section~\ref{sec:lag-irreversibility} we define the lag irreversibility function and we introduce the necessary concepts that we use for the rest of the paper. Section~\ref{sec:methodology} is devoted to describe the methodology employed to obtain the signals, the filtering process, the symbolization of the time series and the methodology to estimate the lag irreversibility function. We give our results in Section~\ref{sec:results}, as well as we test them using the ROC analysis. Finally we give some concluding remarks in Section~\ref{sec:concluding}.

\section{Lag irreversibility} \label{sec:lag-irreversibility}

\subsection{Irreversibility and entropy production}

Let us start by stating some preliminary concepts about irreversibility of stochastic processes. Let $\mathcal{X} = \{X_t \, : \, t \in \mathbb{N}\}$  be a discrete-time stochastic process where the state variables take values on a finite state space $\mathcal{S}$.  Also, we denote by  $\mathbf{X}_t^{t+n} $ a finite random path or trajectory of the process starting at time $t$ and ending a time $t + n$, i.e.,
\begin{eqnarray}
\mathbf{X}_t^{t+n}  = (X_t,X_{t+1}, \dots, X_{t+n}).
\end{eqnarray}
The realizations of the process or the finite trajectories will be denoted by bold lower case letters, i.e., a realization of the random path $\mathbf{X}_t^{t+n}$ will be denoted simply by $\mathbf{a}$, where $\mathbf{a}$ belongs to the set $\mathcal{S}^{n+1}$. If necessary, we will use subscripts and superscripts to emphasize the finite character of the realizations, $\mathbf{a}_0^{n}$.  

We know that the stochastic process $\mathcal{X}$ is reversible if for all $n\in \mathbb{N}$, all $t\in \mathbb{N}_0$ and any finite sequence $\mathbf{a}\in \mathcal{S}^{n+1}$ we have that
\begin{eqnarray}
\mathbb{P}(\mathbf{X}_t^{t+n} = \mathbf{a} ) = \mathbb{P}(\mathbf{X}_t^{t+n} = \overline{\mathbf{a}} ),
\label{eq:def:reversibleP}
\end{eqnarray}
where the overline in $ \mathbf{a} $ denotes the reversed realization of the process, i.e., if $\mathbf{a} = (a_0, a_1, \dots, a_{n-1}, a_{n})$, then,
\begin{eqnarray}
\overline{\mathbf{a}}  = (a_{n},a_{n-1},\dots, a_1,a_0).
\end{eqnarray}

If the identity~\eqref{eq:def:reversibleP} does not hold it is said that the process is irreversible.  In that case the probability of observing a given trajectory $\mathbf{a}$ does not coincide with the probability of observing the reverse trajectory  $\overline{\mathbf{a}}$.

The degree of irreversibility of a process $\mathcal{X}$ is commonly measured by means of the Kullback-Leibler divergence between path distribution $\mathbb{P}(\mathbf{X}_t^{t+n} = \mathbf{a} )$ and  the reversed path distribution $\mathbb{P}(\mathbf{X}_t^{t+n} = \overline{\mathbf{a}} )$:
\begin{eqnarray}
D_{\mathrm{KL}} (\mathbb{P}(\mathbf{X}_t^{t+n} = \mathbf{a} ) \, || \,
\mathbb{P}(\mathbf{X}_t^{t+n} =\overline{\mathbf{a}} ))
:= \sum_{\mathbf{a}\in \mathcal{S}} \mathbb{P}(\mathbf{X}_t^{t+n} = \mathbf{a} )  \log \bigg( \frac{\mathbb{P}(\mathbf{X}_t^{t+n} = \mathbf{a} ) }{ \mathbb{P}(\mathbf{X}_t^{t+n} = \overline{\mathbf{a}} ) } \bigg).
\end{eqnarray}
The previous quantity is related to the entropy production rate, which is an important concept. From the physical point of view, it is not only a measure of the time irreversibility, but it provides information about the nature of the corresponding physical system. For example, it is known that if a given time series comes from a thermodynamic system at equilibrium, then the entropy production rate  of such a process will be zero, and thus, being time-reversible. Otherwise, if such a time series comes from a system out from equilibrium, the series will be clearly irreversible. Despite its importance, from the practical point of view, the Kullback-Leibler divergence between the path and reversed path distributions has it drawbacks while being estimated from a sample trajectory (or finite realization). This is mainly  due to the fact that it is not only necessary to take a limit for $n \to \infty$, but also it is necessary to obtain estimations for the $n$-dimensional marginals $\mathbb{P}(\mathbf{X}_t^{t+n} = \mathbf{a} ) $, for which  their fluctuations might be large when $n$ increases, since the size of the sample is finite. This fact, among others, has motivated the introduction of several techniques to assess the time-irreversibility, something that has been achieved with different success in every case~\cite{Zanin2021Algorithmic}.

\subsection{Lag irreversibility}

As we mentioned above, the main problem in evaluating the degree of irreversibility from a time series is the finiteness of the sample. Some estimators like those based on recurrence times, such as hitting time or  waiting time, require a large sample trajectory. Other estimators might be more efficient and a comparative study can be found in~\cite{Zanin2021Algorithmic}. The  main advantage of the method we introduce here for determining the irreversibility of given time series, is that the number of parameters to be estimated is relatively small compared to other methods. 
This method will be referred to as \textit{lag irreversibility} and the basic idea of the proposed irreversibility index goes as follows. First consider a discrete-time stochastic process $\mathcal{X} = \{X_t \, : \, t \in \mathbb{N}\}$ and let $\tau \in \mathbb{N}$ be a non-negative integer. Assume that such a stochastic process is stationary and consider the random path  $\mathbf{X}_t^{t+\tau} = (X_t, X_{t+1}, \dots, X_{t+\tau})$. As we mentioned above, if the process $\mathcal{X} $ is irreversible, then
\begin{eqnarray}
\mathbb{P}(\mathbf{X}_t^{t+\tau} = \mathbf{a}_0^{\tau}  ) \not= \mathbb{P}(\mathbf{X}_t^{t+\tau} = \overline{\mathbf{a}}_0^{\tau} ).
\label{eq:def:irrev}
\end{eqnarray}
If we sum over $a_1, a_2, \dots a_{\tau-1}$ it is clear that
\begin{eqnarray}
\sum_{a_1} \sum_{a_2} \cdots \sum_{a_{\tau-1}}  \mathbb{P}(\mathbf{X}_t^{t+\tau} = \mathbf{a}_0^{\tau}  ) = \mathbb{P}({X}_t = {a}_0;  X_{t+\tau} = {a}_{\tau} ),
\end{eqnarray}
where $\mathbb{P}({X}_t = {a}_0;  X_{t+\tau} = {a}_{\tau} )$ is the joint probability function for the state variables $X_t$ and $X_{t+\tau}$. Equivalently we have that, 
\begin{eqnarray}
\sum_{a_1} \sum_{a_2} \cdots \sum_{a_{\tau-1}}  \mathbb{P}(\mathbf{X}_t^{t+\tau} = \overline{\mathbf{a}}_0^{\tau}  ) = \mathbb{P}({X}_t = {a}_{\tau};  X_{t+\tau} = {a}_{0} ),
\end{eqnarray}
where $\mathbb{P}({X}_t = {a}_{\tau};  X_{t+\tau} = {a}_{0} )$ is the joint probability function for the state variables $X_t$ and $X_{t+\tau}$, but reversed in time.
The above equations, together with the condition of irreversibility given by Eq.~(\ref{eq:def:irrev}) implies that if the process  $\mathcal{X}$ is irreversible, then the joint distribution for the state variables  $X_t$ and $X_{t+\tau}$ will have the following asymmetry property,
\begin{eqnarray}
\mathbb{P}({X}_t = {a}_0;  X_{t+\tau} = {a}_{\tau}  ) \not= \mathbb{P}({X}_t = {a}_{\tau};  X_{t+\tau} =  {a}_{0} ).
\label{eq:def:pair-wise-irrev-pre}
\end{eqnarray}
The latter motives our definition of \textit{pairwise reversibility}.

\begin{definition}
We say that a stochastic process $\mathcal{X} = \{X_t \, : \, t \in \mathbb{N}\}$ is pairwise reversible if for all $\tau\in \mathbb{N} $, we have that
\begin{eqnarray}
\mathbb{P}({X}_t = {a}_0;  X_{t+\tau} = {a}_{\tau}  )  = \mathbb{P}({X}_t = {a}_{\tau};  X_{t+\tau} = {a}_{0} ).
\end{eqnarray}
Otherwise, we  say that the process is pairwise irreversible.

\end{definition}

To check pairwise irreversibility we need to estimate the joint probability function $\mathbb{P}({X}_t = {a}_0;  X_{t+\tau} = {a}_{\tau}  )$ for every $\tau$, a task that is computationally less expensive than estimating the joint probability function of the whole trajectory. A natural way to evaluate the pairwise irreversibility is to compute the Kullback-Leibler divergence of the joint probability $\mathbb{P}({X}_t = {a}_0;  X_{t+\tau} = {a}_{\tau}  )$ with respect to $\mathbb{P}({X}_t = {a}_{\tau};  X_{t+\tau} = {a}_0  )$, which leads to the following definition of the lag irreversibility function.

\begin{definition}
We define the lag irreversibility (LI) function, $L(\tau)$,  as the Kullback-Leibler divergence of the joint probability $\mathbb{P}({X}_t = {a}_0;  X_{t+\tau} = {a}_{\tau}  )$ with respect to $\mathbb{P}({X}_t = {a}_{\tau};  X_{t+\tau} = {a}_0  )$, i.e.,
\begin{eqnarray} 
L(\tau) &:=& D_{KL} \big( \mathbb{P}({X}_t = {a}_0;  X_{t+\tau} = {a}_{\tau}  ) 
\, || \,  \mathbb{P}({X}_t = {a}_{\tau};  X_{t+\tau} = {a}_0  ) \big) 
\nonumber
\\
&=&
\sum_{a_0} \sum_{a_{\tau}}  \mathbb{P}({X}_t = {a}_0;  X_{t+\tau} = {a}_{\tau}  ) 
\log\bigg( \frac{\mathbb{P}({X}_t = {a}_0;  X_{t+\tau} = {a}_{\tau}  ) }
{ \mathbb{P}({X}_t = {a}_{\tau};  X_{t+\tau} = {a}_0  ) \big) }\bigg).  \quad \quad
\label{LagirreversibilityEqu}
\end{eqnarray} 

\end{definition}

It is important to remark that we named \emph{lag irreversibility function}  to the Kullback-Leibler divergence $L(\tau)$   because, in some sense, we are measuring the asymmetry of the joint  probability function of the state variable  $X_t$ and the state variable lagged a time $\tau$, i.e., $X_{t+\tau}$. This asymmetry, as we argued above, gives information about the irreversibility of the process throughout the estimation of $L(\tau)$ for every $\tau$.

At this point, it is instructive to compute analytically the  lag irreversibility function  $L(\tau)$ in a specific Markov chain model.
\begin{example}
Let  $\mathcal{X} = \{X_t \, : \, t \in \mathbb{N}\}$  be a three-states stationary Markov chain with state space $\mathcal{S} = \{1,2,3\}$ and stochastic matrix $\mathbf{Q}$ given by
\begin{eqnarray}
\mathbf{Q} =
\left[
  \begin{array}{ccc}
   0      & p      & 1-p  \\
   1-p   & 0      &   p   \\
      p   & 1-p   &   0   
  \end{array} 
\right].
\end{eqnarray}

%
\begin{figure}[t]
\begin{center}
\scalebox{0.4}{\includegraphics{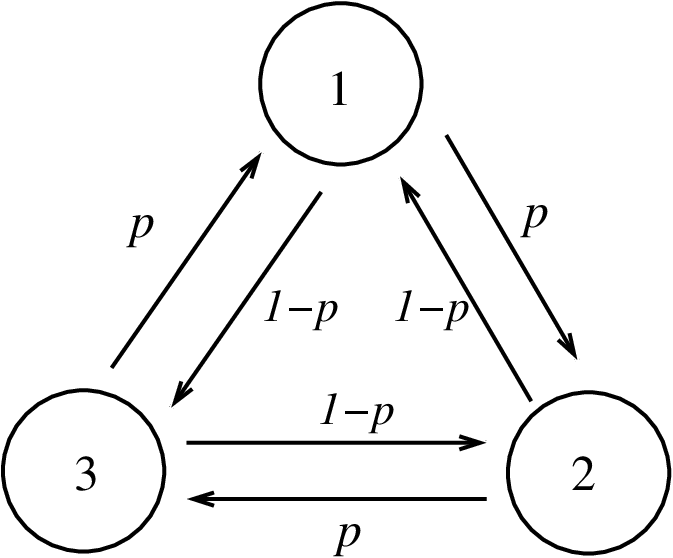}}
\end{center}
     \caption{
      The three-states Markov chain model. This Markov chain is known to be reversible if $p= 1/2$. For parameter values lower or larger than $p = 1/2$ the Markov chain is irreversible, and the irreversibility get larger as $p$ gets away the value $p = 1/2$. 
            }
\label{fig:fig01}
\end{figure}
%
\noindent
It is known that for a Markov chain, the joint probability function $\mathbb{P}({X}_t = {a}_0;  X_{t+\tau} = {a}_{\tau}  ) $ can be written as
\begin{eqnarray}
\mathbb{P}(  {X}_t = {a}  ; X_{t+\tau} = b ) = \pi_a \left(  \mathbf{Q}^\tau \right)_{a,b},
\end{eqnarray}
where $a,b \in \mathcal{S}$ and $\pi_a $ is the $a$-th element of the stationary probability vector $\boldsymbol{\pi}$, i.e., the probability vector  satisfying the stationary equation $ \boldsymbol{\pi} = \boldsymbol{\pi} \mathbf{Q} $. 

It is not hard to see that $L(\tau)$ can be written in this case as,
\begin{eqnarray}
L(\tau) = \sum_{a\in \mathcal{S}}  \sum_{b\in \mathcal{S}}  \pi_a \left(  \mathbf{Q}^\tau \right)_{a,b}
\log\bigg(  \frac{ \left(  \mathbf{Q}^\tau \right)_{a,b}}{ \left(  \mathbf{Q}^\tau \right)_{b,a}}   \bigg).
\label{eq:Ltau_1step-MC}
\end{eqnarray}
%
\begin{figure}[ht]
\begin{center}
\scalebox{0.5}{\includegraphics{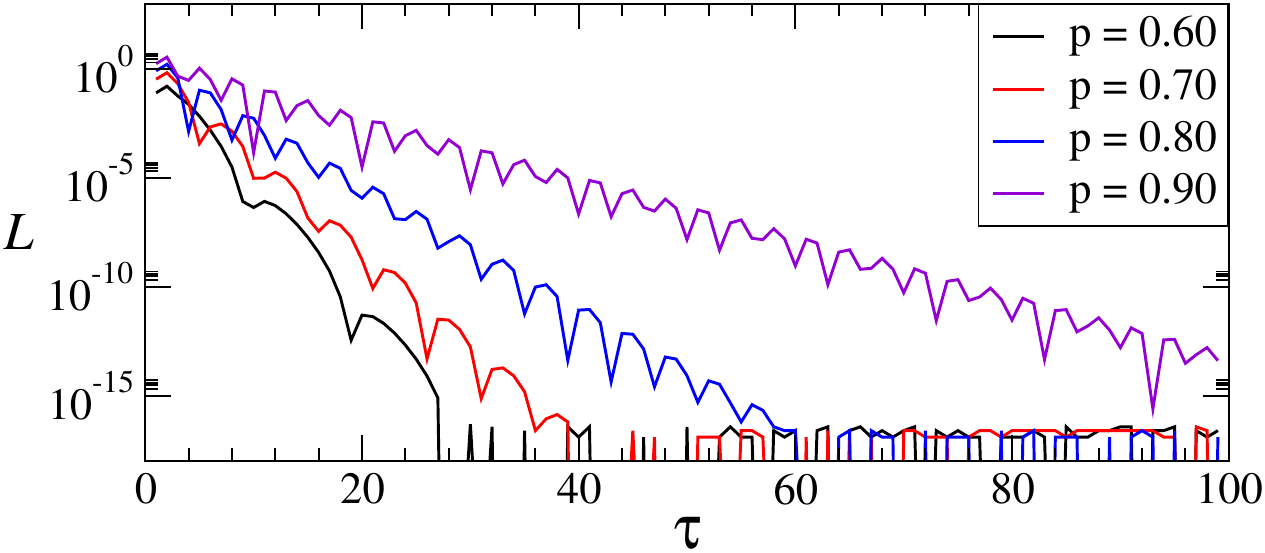}}
\end{center}
     \caption{
      Lag-irreversibility function for the three-states Markov chain. We show the LI functions for $p=0.60$ (solid black line), $p=0.70$ (solid red line), $p=0.80$ (solid blue line), and $p=0.90$ (solid violet line). Observe that the LI function gets larger as $p$ increases. Also is important to notice that all the LI functions decay exponentially fast and the larger $p$ the slower decaying rate. 
                  }
\label{fig:fig02} 
\end{figure}
%
Eq.~(\ref{eq:Ltau_1step-MC}) is actually the expression for the lag irreversibility function for any one-step Markov chain. This expression can be straightforwardly   evaluated  for the three-states Markov chain introduced above. In Fig.~\ref{fig:fig02} we can appreciate the behavior of the LI function for several values of the parameter $p$. All the parameter values we use to compute the lag irreversibility function for this Markov model correspond to the case in which the process is irreversible. Actually the only parameter value for which the three-states Markov chain is reversible corresponds to $p=1/2$. It is not hard to see that the LI function, $L(\tau)$, is zero for all $\tau \in \mathbb{N}$ in the case $p=1/2$. This is because in this case the stochastic matrix is symmetric, implying that 
\[
 \left( \mathbf{Q}^\tau \right)_{a,b} =   \left( \mathbf{Q}^\tau \right)_{b,a},
\] 
making  $L(\tau) = 0$ for all $\tau$. For other values of the parameter $p$, the LI function is positive and increases as $p$ gets away from $p=1/2$.
In Fig.~\ref{fig:fig02} we show the LI functions for $p=0.60$,  $p=0.70$,  $p=0.80$, and  $p=0.90$. As we see, the LI function gets larger as the parameter $p$ increases. Also, as $p$ increases, the entropy production also increases,  which is consistent with the fact that the  LI function can be used as a measure of the  irreversibility of the process.  We should notice that the larger entropy production rate, the lower decaying rate in the LI function, which allows to conclude that the decay rate of the LI function might be a measure (or at least an indicator) of the irreversibility as well.

\end{example}

As we saw in the preceding example, we have that the LI function $L(\tau)$ decays as $\tau$ increases. This is actually a consequence of the mixing property of the Markov chain, i.e., as $\tau$ gets larger, the random variables  $X_t$ and $X_{t+\tau}$ become independent,  that is known as the \textit{decay of correlations}. If the random variables  $X_t$ and $X_{t+\tau}$ become independent, then the joint probability function $\mathbb{P}(  {X}_t = {a}  ; X_{t+\tau} = b )$ becomes symmetric, i.e., 
\begin{eqnarray}
\big| \mathbb{P}(  {X}_t = {a}  ; X_{t+\tau} = b )- \mathbb{P}(  {X}_t = {b}  ; X_{t+\tau} = a) \big| \to 0,
\end{eqnarray}
as $\tau \to \infty$.  This implies that, if a given process comply with the mixing property, we will have that 
\[
\lim_{\tau \to \infty } L(\tau)  = 0.
\]

\subsection{Lag irreversibility estimator}
The empirical lag irreversibility estimator is defined by the empirical joint probability of the state variables $X_t$ and $X_{t+\tau}$, both in the original sequence and in the time-reversed sequence. We assumme that these variables are produced by an unknown processes, but that they can be estimated directly from the trajectories. Let $x=x_1,x_2,…,x_n$ be a realization of the stationary process $\mathcal{X} = \{X_t \, : \, t \in \mathbb{N}\}$, with state space $\mathcal{S}=\{a_1,a_2,…,a_N\}$. We say that process $\mathcal{X}$ at time $i$ is in state $a_j$ if $x_i=a_j$. The joint probability functions $\mathbb{P}(  {X}_t = {a_0}  ; X_{t+\tau} = a_{\tau} )$ and $\mathbb{P}(  {X}_t = {a_{\tau}}  ; X_{t+\tau} = a_0)$, can be estimated by means of
\begin{subequations}
\begin{equation}
\mathbb{\widehat{P}}(  {X}_t = {a_0}  ; X_{t+\tau} = a_{\tau} )=\frac{1}{n-\tau} \sum_{t=1}^{n-\tau} \chi_{a_{0}}(x_{t}) \chi_{a_{\tau}}( x_{t+\tau}),
\label{jointprobabilityEstimation}
\end{equation}
\textit{and}
\begin{equation}
\mathbb{\widehat{P}}(  {X}_t = {a_{\tau}}  ; X_{t+\tau} = a_0)=\frac{1}{n-\tau} \sum_{t=1}^{n-\tau} \chi_{a_{\tau}}(x_{t}) \chi_{x_{a_{0}}}(x_{t+\tau}),
\label{jointprobabilityEstimationReversed}
\end{equation}
\end{subequations}
respectively, where $ \chi_{a}(\cdot)$ is the indicator function for the state $a$. Equations~(\ref{jointprobabilityEstimation}) and~(\ref{jointprobabilityEstimationReversed}) give us the empirical joint probability that we use to estimate $\widehat{L}(\tau)$, by directly substitute them in Equation~(\ref{LagirreversibilityEqu}).

\section{Methodology} \label{sec:methodology}


As we mentioned above, the main purpose of this work is to use the time irreversibility as a property for discriminating among four different groups of patients, namely, Healthy Young (HY), Healthy Elderly (HE), Congestive Heart Failure (CHF) and Atrial Fibrillation (AF). However, quantifying the degree of irreversibility from ECG recordings is neither a direct nor a trivial task, so we propose a methodology consisting of six stages. (1) retrieving and selecting data (ECG recordings) from databases, (2) cleaning ECG recordings and computing variability signals from ECGs (see definitions below), (3) symbolic encoding of variability signals, (4) designing estimation method, (5) assessing irreversibility for classification and (6) evaluating classification through ROC analysis. Next, we give a brief description of stages (1)-(4) which are the preliminary steps before presenting of our main results, stages (5) and (6), which are fully described in Section~\ref{sec:results} below.

\subsection{Retrieving and selecting data}

Datasets were obtained from the {\em PhysioBank} database~\cite{PhysioBank2000}, by means of the open-source WFDB Software Package~\cite{1ToolBox2014}. Healthy Young (HY) and Healthy Elderly (HE) groups were selected from {\em FANTASIA} database~\cite{iyengar1996age}, while  Congestive Heart Failure (CHF) and Atrial Fibrillation (AF) groups were selected from the {\em BIDM Congestive Heart Failure}~\cite{baim1986survival} and {\em MIT-BIH Atrial Fibrillation}~\cite{moody1983new} databases, respectively.

According to the database, Healthy Young group, as well as Healthy Elderly group, are comprised of 20 electrocardiograms (ECGs) each, which were acquired from patients at supine rest. The sampling frequency is $250$ Hz and every ECG record is 120 minutes long. It is also important to mention that the HY patient group, from which these ECGs were obtained, is comprised of 10 men and 10 women aged between 21 and 34. On the other hand, HE patient group is also comprised of 10 men and 10 women, but aged between 68 and 85. 

For the CHF group, the databases provide 15 ECG recordings from which, 11 corresponds to men aged from 22 to 71 and 4 corresponds to women aged from 54 to 63. Every ECG record in this group is 20 hours long with a sampling frequency of $250$ Hz. For the AF group the databases provides 25 ECG records of adult subjects~\footnote{No mention about the patient age is found in the database.} and each recording is 10 hours long with a sampling frequency of $250$ Hz. ECG recordings for AF and CHF groups were obtained by means of ambulatory electrocardiography.

At this point, it is important to stress that, although HY, HE and AF databases have 20 or more electrocardiographic recordings, CHF database contains only 15 ECG samples. Therefore, we only considered 15 ECG records for each group. Also, in order to have homogeneous samples, we consider $5\,000$ heartbeats from ECGs to carry out all the estimates. This choice is due to the fact that this is the least common number of heartbeats within these selected databases.

\subsection{Cleaning ECG records and variability signals}

\begin{figure}
  \centering
  \includegraphics[width=1.0\linewidth]{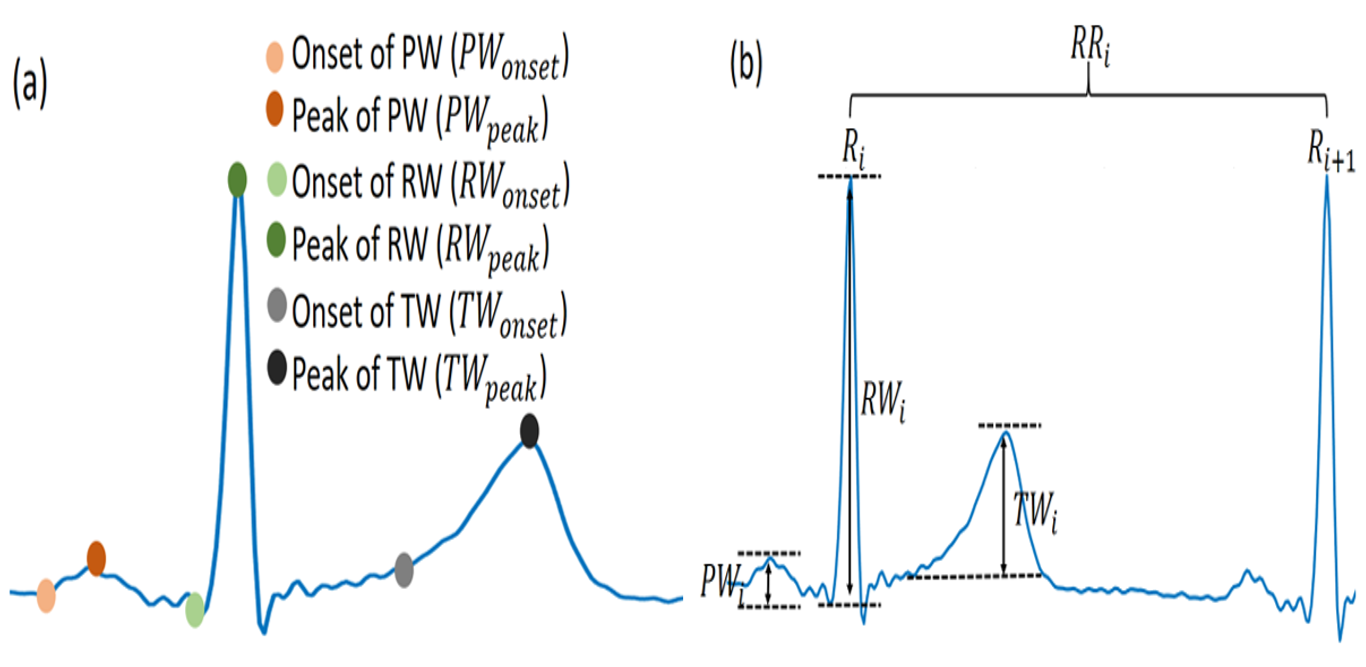}  
  \caption{Components of the ECG signal. The amplitudes of waves P, R and T, at the $i$-th heartbeat, are obtained by means of $PW_i=PW_{peak_i}-PW_{onset_i}$, $RW_i=RW_{peak_i}-RW_{onset_i}$ and $TW_i=TW_{peak_i}-TW_{onset_i}$, respectively.}
\label{ECGsignal_components}
\end{figure}

The ECG recordings retrieved from databases are not completely clean, and might possess some artifacts that could not allow the appropriate identification of P, Q, R, S and T waves in the ECG. To deal with this issue, we made use of  {\em NeuroKit2} package~\cite{Makowski2021neurokit}, which is The Python Toolbox for Neurophysiological Signal Processing. This tool allows to clean ECG signals and detect the components, i.e., the P, Q, R, S, and T waves. In particular, we can automatically clean the ECG and locate the onsets and peaks of the P, R and T waves (see Figure~\ref{ECGsignal_components}(a)), using the \verb|ecg_process|, \verb|ecg_peaks| and \verb|ecg_delineate| commands.

Next, once we obtained the  onsets and peaks of  P, R and T waves, we compute the wave amplitude by means of the difference between the voltage at its onset point and its peak. In other words, the wave amplitude is defined as $W_{peak}-W_{onset}$, where $W_{onset}$ is the voltage at the starting point of the wave and $W_{peak}$ is the voltage that reaches the peak of the wave. For example, for the $i$-th heartbeat, the P-wave amplitude is given by $PW_i=PW_{peak_i}-PW_{onset_i}$; we repeat the procedure analogously for the R and T waves (see Figure \ref{ECGsignal_components}(b)). Regarding the RR intervals, whose length is called \emph{heart rate variability} (HRV), it is possible to obtain them using the \verb|hrv| function of {\em Neurokit}. 

This definition of wave amplitude gives us four time series, namely, $PW$, $RW$, $TW$  and the heart rate variability RR. These quantities are those that we refer to as \emph{variability signals}, which are discrete time series of the form $x=(x_1,x_2,\dots,x_{m})$, where $m=5\,000$, is the number of heartbeats contained in data. To eliminate spurious non-stationary characteristics, we consider the differences of consecutive entries of each time series, i.e., we use the vector $y=(y_1,y_2,\dots,y_{m-1})$, where $y_{i}=x_{i+1}-x_i$, for $i=1,2,\dots,m-1$. Specifically, for the RR intervals, P-wave, R-wave, and T-wave, their respective difference time series are given by $\Delta RR=\{ \Delta RR_i:\Delta RR_i=RR_{i+1}-RR_{i}, i=1,2,\dots,m-1\}$, $\Delta PW=\{ \Delta PW_i:\Delta PW_i=PW_{i+1}-PW_{i}, i=1,2,\dots,m-1\}$, $\Delta RW=\{ \Delta RW_i:\Delta RW_i=RW_{i+1}-RW_{i}, i=1,2,\dots,m-1\}$ and $\Delta TW=\{ \Delta TW_i:\Delta TW_i=TW_{i+1}-TW_{i}, i=1,2,\dots,m-1\}$.

\subsection{Symbolic encoding} \label{symbolicEnconding}
Once we have the time series of the  $\Delta RR$, $\Delta PW$, $\Delta RW$ and $\Delta TW$ signals, we proceed to encode them by partitioning the state space of the time series. An intuitive encoding technique was proposed in \cite{Daw2000Symbolic}, which considers a uniform partition of the state space. However, this uniform partition does not take into account the typical fluctuations of the signal. In order to capture these fluctuations inherent to physiological processes, in~\cite{Nazul}, the authors proposed different partitions composed by elements with non-uniform sizes. There, it is considered a ``center cell'' around the mean $\mu$ value of the time series, and the size of the element of the partition is defined using the standard deviation $\sigma$ of the data and fitted by means of a parameter $\gamma$. After that, the authors make an exploratory study in order to select a suitable value for $\gamma$ in the sense that the obtained results permit a better discrimination among the groups of patients.

\begin{figure}[!htb]
\begin{center}
\begin{tikzpicture}[scale=1]
\draw [<->, thick] (0,-0.5) -- (0,3.5);
\draw[ white,fill= gray!20] (0,1) rectangle (4,2);
\draw [decorate,
    decoration = { brace,
        amplitude=5pt, mirror}, thick] (4.1, 1) --  (4.1,2);
\draw (4.5,1.5) node[right] {$2$};
\draw [-] (-0.5,1.5) -- (4, 1.5);
\draw (-0.3,1.2) node[left] {$\mu$};
\draw [|-|] (0,1) -- (0,2);

\draw [decorate,
    decoration = { brace,
        amplitude=5pt, mirror}, thick] (4.1, 2) --  (4.1,3.5);
\draw (4.5,2.7) node[right] {$3$};

\draw [decorate,
    decoration = { brace,
        amplitude=5pt, mirror}, thick] (4.1, -0.5) --  (4.1, 1);
\draw (4.5,0.25) node[right] {$1$};

 \draw (0.4, 0.3) circle [radius=0.05];
 \draw (1, 1.7) circle [radius=0.05];
 \draw (1.5, 0.8) circle [radius=0.05];
 \draw (2, 2.8) circle [radius=0.05];
 \draw (2.5, 1.1) circle [radius=0.05];
\draw (3, 1.9) circle [radius=0.05];
\draw (3.5, 0.1) circle [radius=0.05];

\draw [dotted] (0.4,0.4) -- (1, 1.7);
\draw [dotted] (1,1.7) -- (1.45,0.85);
\draw [dotted] (1.5,0.85) -- (2, 2.8);
\draw [dotted] (2, 2.75) -- (2.5,1.1);
\draw [dotted] (2.55,1.15) -- (3,1.9);
\draw [dotted] (3,1.8) -- (3.5, 0.15);

\draw [->] (0.38,0.15) -- (0.4, -0.7);
\draw (0.38, -0.8) node[below] {$1$};

\draw [->] (1,1.4) -- (1, -0.7);
\draw (1, -0.8) node[below] {$2$};

\draw [->] (1.5,0.65) -- (1.5, -0.7);
\draw (1.5, -0.8) node[below] {$1$};

\draw [->] (2, 2.6) -- (2, -0.7);
\draw (2, -0.8) node[below] {$3$};

\draw (2.6, -0.8) node[below] {$\cdots$};
\draw (-1,1.5) node[left] {$\mu \pm \gamma \sigma$};
\draw (-1,2.7) node[left] {$\geq \mu + \gamma \sigma$};
\draw (-1,0.2) node[left] {$\leq \mu - \gamma \sigma$};

\end{tikzpicture}
\end{center}
\caption{Schematic diagram of the encoding  method for a single signal.}
\label{fig:encoding}
\end{figure}
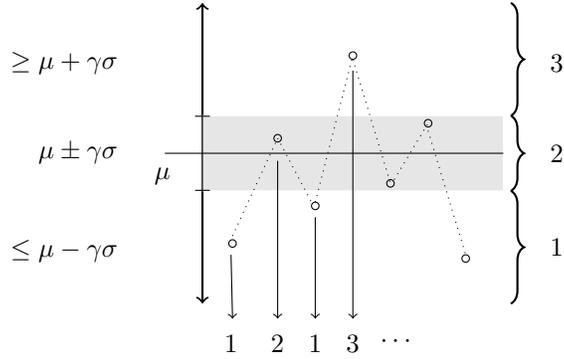

Here, we simplify the method for defining a useful partition from~\cite{Nazul}. We consider the simplest partition of the state space that allows a good estimation of the irreversibility properties, which is a ternary partition. And in oder to draw information from the time series, we perform a {\em single signal encoding} and a {\em joint signal encoding}, considering non-uniform sizes of the partitions. This enables us, on the one hand, to simplify the method for obtaining the partition; and, on the other hand, to obtain more information about the irreversibility properties of the data sequences. For the {\em single signal encoding} (see Figure~\ref{fig:encoding} for a schematic representation), given a time series of differences $y=(y_1,y_2,…,y_{m-1})$, with mean $\mu$ and standard deviation $\sigma$, its corresponding symbolic sequence can be obtained by using the rule:
\[
s_i :=
\begin{cases}
1 & \mbox{ if }\ y_i \leq \mu-\gamma \sigma, \\
2 & \mbox{ if }\ \mu-\gamma \sigma <  y_i <  \mu+\gamma \sigma, \\
3 & \mbox{ if }\ y_i \geq \mu+\gamma \sigma.\\
\end{cases}
\]


For the case of {\em joint signal encoding}, given two time series of differences $y^1=(y_1^1,y_2^1,…,y_{m-1}^1)$ (with parameters $\mu_1$ and $\sigma_1$) and $y^2=(y_1^2,y_2^2,…,y_{m-1}^2)$ (with parameters $\mu_2$ and $\sigma_2$), their joint symbolic sequence can be obtained through the following rule:
\[ s_i= \left\{ \begin{array}{lccc}
1\quad \mbox{if}  & y_{i}^{1} \geq \mu_1+\gamma \sigma_1 & \wedge & y_{i}^{2} \geq \mu_2+\gamma \sigma_2, \\
2\quad \mbox{if}  & y_{i}^{1} \geq \mu_1+\gamma \sigma_1 & \wedge & \mu_2-\gamma \sigma_2 <  y_{i}^{2} < \mu_2+\gamma \sigma_2, \\
3\quad \mbox{if}  & y_{i}^{1} \geq \mu_1+\gamma \sigma_1 & \wedge & y_{i}^{2} \leq \mu_2-\gamma \sigma_2, \\
4\quad \mbox{if}  & \mu_1-\gamma \sigma_1  < y_{i}^{1} < \mu_1+\gamma \sigma_1 & \wedge & y_{i}^{2} \geq \mu_2+\gamma \sigma_2, \\
5\quad \mbox{if}  & \mu_1-\gamma \sigma_1  < y_{i}^{1} < \mu_1+\gamma \sigma_1 & \wedge & \mu_2-\gamma \sigma_2 <  y_{i}^{2} < \mu_2+\gamma \sigma_2, \\
6\quad \mbox{if}  & \mu_1-\gamma \sigma_1  < y_{i}^{1} < \mu_1+\gamma \sigma_1 & \wedge &  y_{i}^{2} \leq \mu_2-\gamma \sigma_2, \\
7\quad \mbox{if}  & y_{i}^{1} \leq \mu_1-\gamma \sigma_1 & \wedge & y_{i}^{2} \geq \mu_2+\gamma \sigma_2, \\
8\quad \mbox{if}  & y_{i}^{1} \leq \mu_1-\gamma \sigma_1 & \wedge & \mu_2-\gamma \sigma_2 <  y_{i}^{2} < \mu_2+\gamma \sigma_2, \\
9\quad \mbox{if}  & y_{i}^{1} \leq \mu_1-\gamma \sigma_1 & \wedge & y_{i}^{2} \leq \mu_2-\gamma \sigma_2. \\
                  \end{array}
\right.
\]

\subsection{Estimation method} \label{estimationprocedure}

Once we have the symbolic sequences of the data, we proceed to build the estimator. First, we estimate the joint probability functions $\mathbb{P}({X}_t = {a}_0;  X_{t+\tau} = {a}_{\tau} )$ and $\mathbb{P}({X}_t = {a}_{\tau};  X_{t+\tau} = {a}_{0})$, for $\tau=1,…20$, using equations \eqref{jointprobabilityEstimation} and \eqref{jointprobabilityEstimationReversed}, respectively. Next, we estimate the {\em lag irreversibility function}, $\widehat{L}(\tau)$, by directly plugging the estimated joint probability functions into equation \eqref{LagirreversibilityEqu}.

\section{Results} \label{sec:results}
%

In this section we estimate the lag irreversibility function for single and joint variability signals. We do that by means of an empirical estimation of the joint probabilities, as described in Section~\ref{estimationprocedure}. For the symbolization process, we consider a ternary partition of the state space,  making an exploratory study for finding the best parameter value $\gamma$.

\subsection{Fixing the value of $\gamma$}

\begin{figure}[t]
\begin{center}
\scalebox{0.4}{\includegraphics{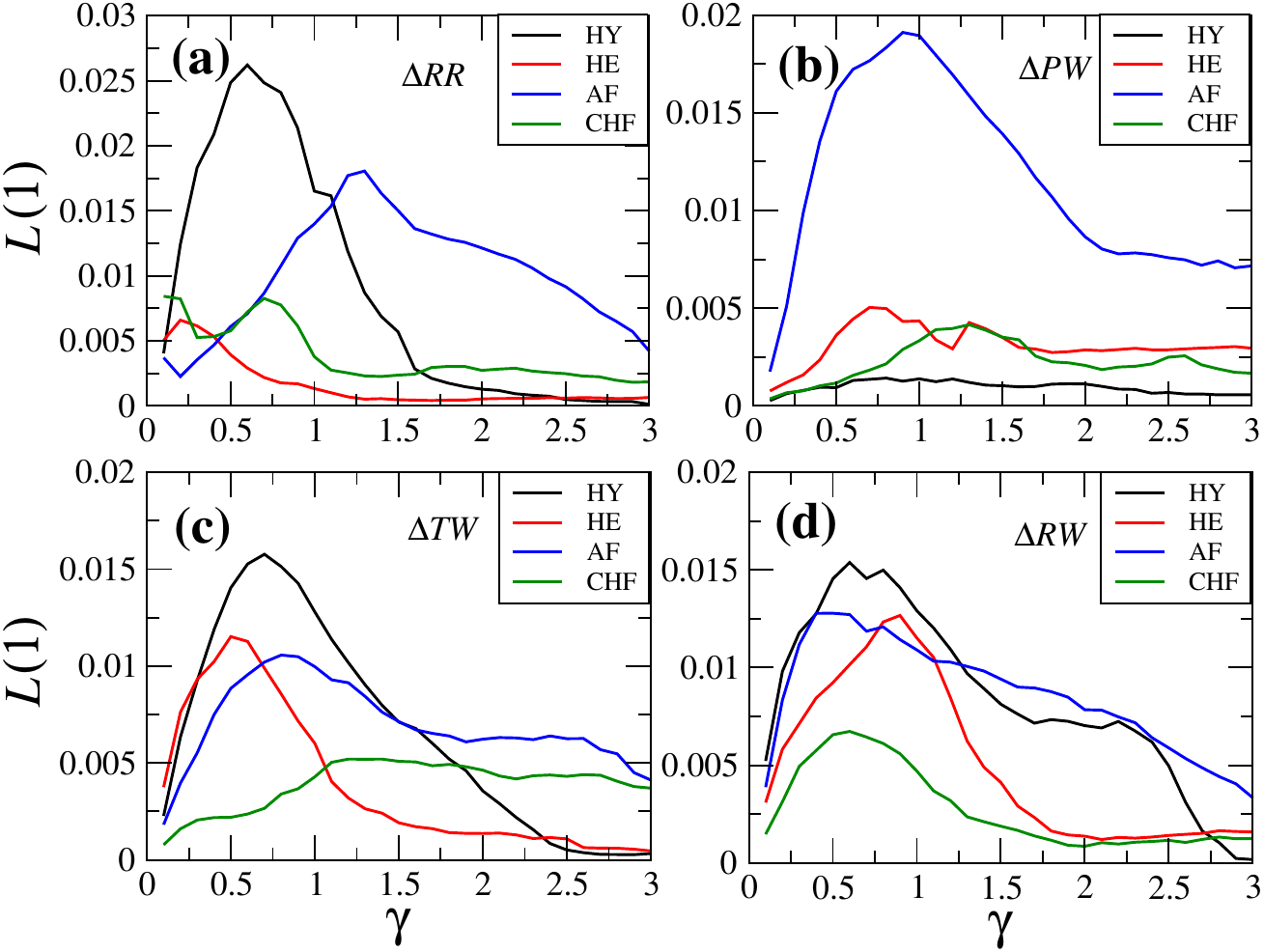}}
\end{center}
     \caption{
       Lag-irreversibility function of single variability signals, for different values of $\gamma$ and for $\tau=1$.
            }
\label{gamma_singleSignals}
\end{figure}

In Figure~\ref{gamma_singleSignals} we show the values of the LI  function for single variability signals, for different values of $\gamma$ and $\tau=1$. In this figure we can see that the heart variability signal ($\Delta RR$) exhibits a higher value of LI for the HY group and it decreases with aging and disease. Furthermore, we see from the latter that a suitable choice for the parameter is  $\gamma=3/10$, in the sense that it allows a better discrimination between groups, i.e., for  healthy young subjects the LI value ($\approx0.01832$) is more than three times higher than the LI value for CHF ($\approx0.00526$) and about five times higher than the LI value for AF group ($\approx0.00354$). For healthy elderly subjects its LI value ($\approx0.00614$) is approximately $16\%$ higher than for CHF ($\approx0.00526$) and $73\%$ higher than for AF ($\approx0.00354$). Finally, the LI value in AF is roughly $48\%$ higher respect the CHF group. From these results we suggest that $\gamma=3/10$ is a suitable choice to estimate $\widehat{L}(\tau)$ as a function of $\tau$ (see Section~\ref{LI_singleVariability} below).

%
%

\begin{figure}[t]
\begin{center}
\scalebox{0.5}{\includegraphics{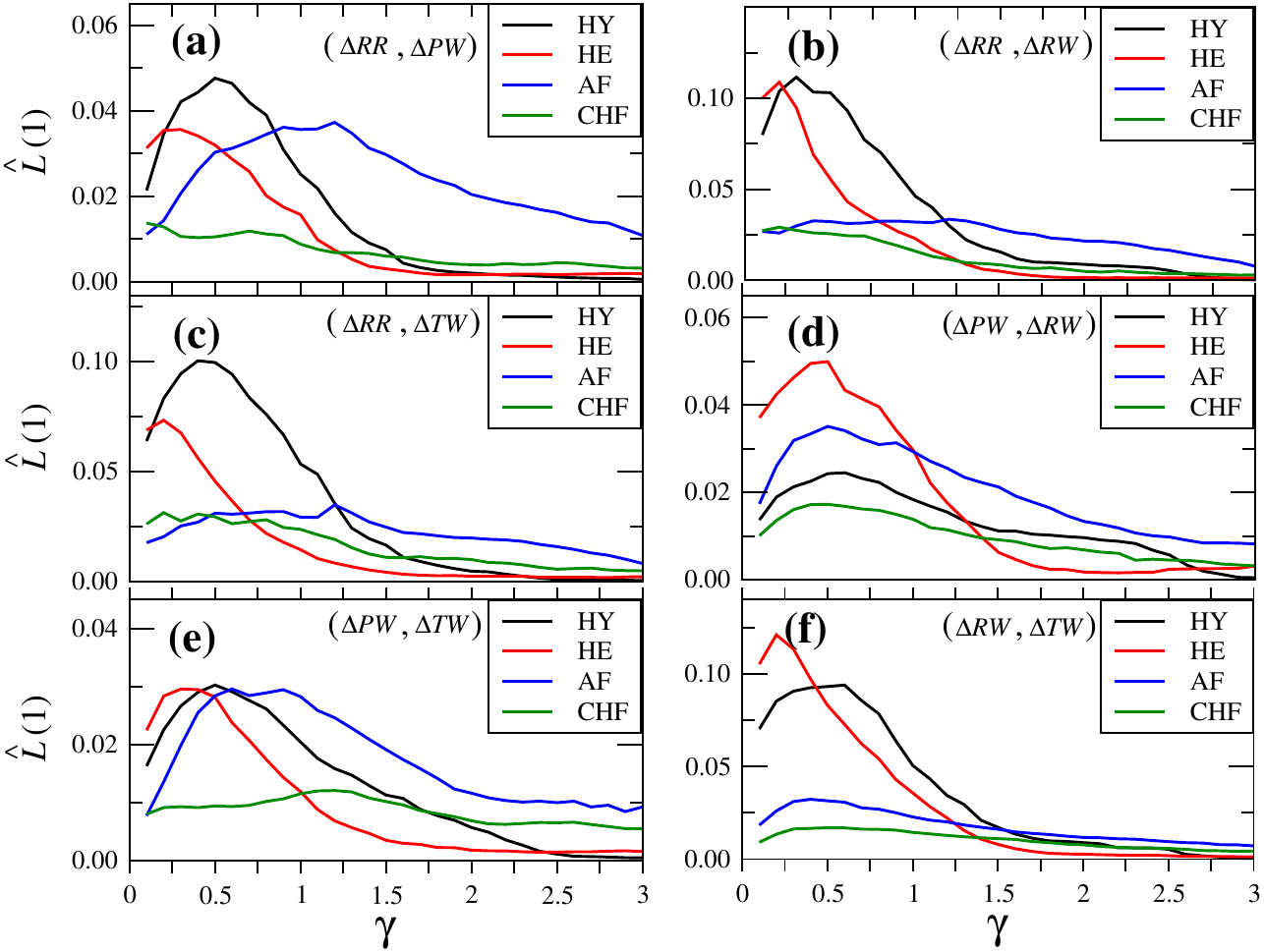}}
\end{center}
     \caption{
       Lag-irreversibility function of joint variability signals, for different values of $\gamma$ and for $\tau=1$.
            }
\label{gamma_jointSignals}
\end{figure}

The main advantage of the lag irreversibility function is found in the case of the joint variability signals. We carry out the same exploratory study, that is, we estimate LI function for $\tau=1$ and different values of $\gamma$, which we show in Figure \ref{gamma_jointSignals}. We can see that joint coding cases involving heart rate variability ($(\Delta RR,\Delta PW), (\Delta RR,\Delta RW),(\Delta RR,\Delta TW)$) allow better and consistent discrimination between groups of healthy patients from those with some adverse health condition. For example, for the case $(\Delta RR,\Delta PW)$, the LI value of the healthy young group is $\approx0.0421$, which is four times greater than that for the CHF group ($\approx0.0106$) and two times greater than the value for the AF group ($\approx0.0207$), while the LI value of the group HE ($\approx0.0356$) is three times greater than that for the CHF group and two times greater than the value for the AF group. Similar results are obtained for $(\Delta RR,\Delta RW),(\Delta RR,\Delta TW)$. All these results were obtained using $\gamma=3/10$ for the three cases (see Table~\ref{Table-joint-DeltaRR} below). Accordingly, we will use that value to estimate $\widehat{L}(\tau)$ as a function of $\tau$, in Section \ref{LI_jointVariability}. 

%
%

\begin{table}[h]
\begin{center}
\begin{tabular}{c|c|c|c|}

\cline{2-4}
 & $\Delta$RR,$\Delta$PW & $\Delta$RR,$\Delta$RW& $\Delta$RR,$\Delta$TW\\
\hline
\multicolumn{1}{|c|}{HY }& 0.04217 & 0.11155 & 0.09455\\
\hline
\multicolumn{1}{|c|}{HE }& 0.03566 & 0.09478 & 0.06762 \\
\hline
\multicolumn{1}{|c|}{CHF}& 0.01063 & 0.02748 & 0.02745 \\
\hline
\multicolumn{1}{|c|}{AF}& 0.02075 & 0.02993 & 0.02519 \\
\hline
\multicolumn{4}{|c|}{$\widehat{L}(\tau=1)$ and $\gamma=3/10$}\\
\hline
\end{tabular}
\caption{ $\widehat{L}(1)$ values for joint variables ($\Delta$RR,$\Delta$PW), ($\Delta$RR,$\Delta$RW) and ($\Delta$RR,$\Delta$TW) using parameter $\gamma=3/10$.}
\label{Table-joint-DeltaRR}
\end{center}
\end{table}

\subsection{Lag-irreversibility of single variability signals} \label{LI_singleVariability}
\begin{figure}[t]
\begin{center}
\scalebox{0.5}{\includegraphics{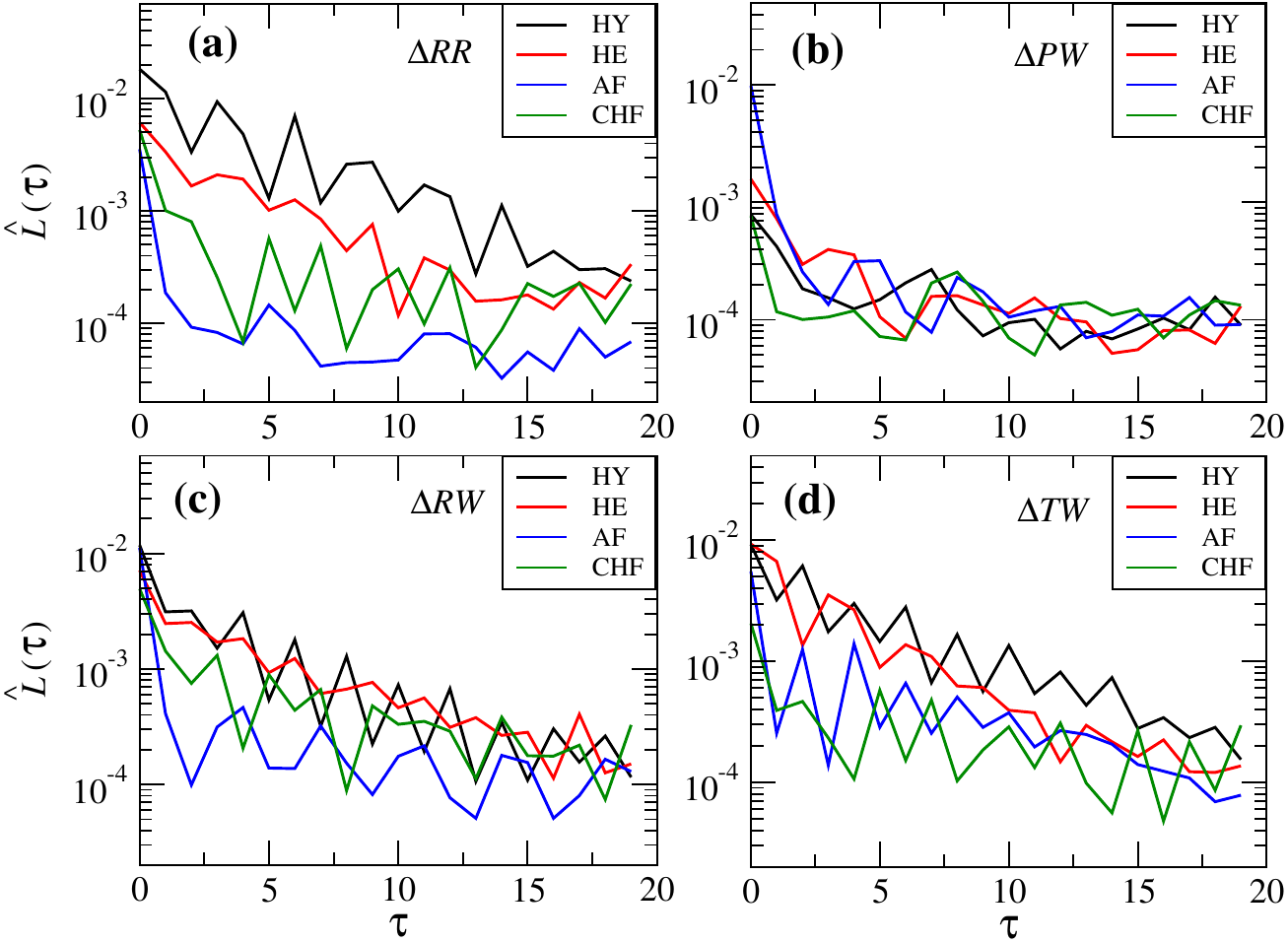}}
\end{center}
     \caption{
       Lag-irreversibility function for single variability signals. We show the estimated LI functions, $\hat L (\tau)$ for (a) the heart rate variability, HRV, 
(b) the P-wave variability,  $\Delta$PV,  (c)  the R-wave variability, $\Delta$RV, and (d) 
 the T-wave variability, $\Delta$TV.
            }
\label{fig:fig01-R}
\end{figure}

In Fig.~\ref{fig:fig01-R} we can see the LI function of all the single variability signals as a function of $\tau$. First of all, we can see in Fig.~\ref{fig:fig01-R}(a) that the LI function for $\Delta RR$ signal for the HY group is larger than for the other groups. The latter means that LI function for the heart rate variability signal might be used as a tool for discriminating between healthy young individuals from the other individuals under study. On the other hand, in Fig.~\ref{fig:fig01-R} (b), (c) and (d), we observe that the LI functions for  $\Delta PW$, $\Delta RW$ and $\Delta TW$ do not allow us to reach the same conclusion. This is because the LI functions look very similar to each other among the different health conditions of individuals.

%
%

\subsection{Lag-irreversibility of joint variability signals}\label{LI_jointVariability}

Provided that the LI function for a single variability signal gives concluding results only for the $\Delta RR$ signal, we go further and consider a joint variability signal study. We expect that the data sequences contain physiological information that can be drawn from the different waves and variability signals. Moreover it is expected that these variability signals are not necessarily independent. This is because all the variability signals are, in some sense, different stages of the heartbeat of the same patient. Thus, the joint distribution  carry some information of the dynamic process that cannot be obtained by solely looking at a single variability signal, such as the heart rate variability or a given wave amplitude variability.

\begin{figure}[t]
\begin{center}
\scalebox{0.55}{\includegraphics{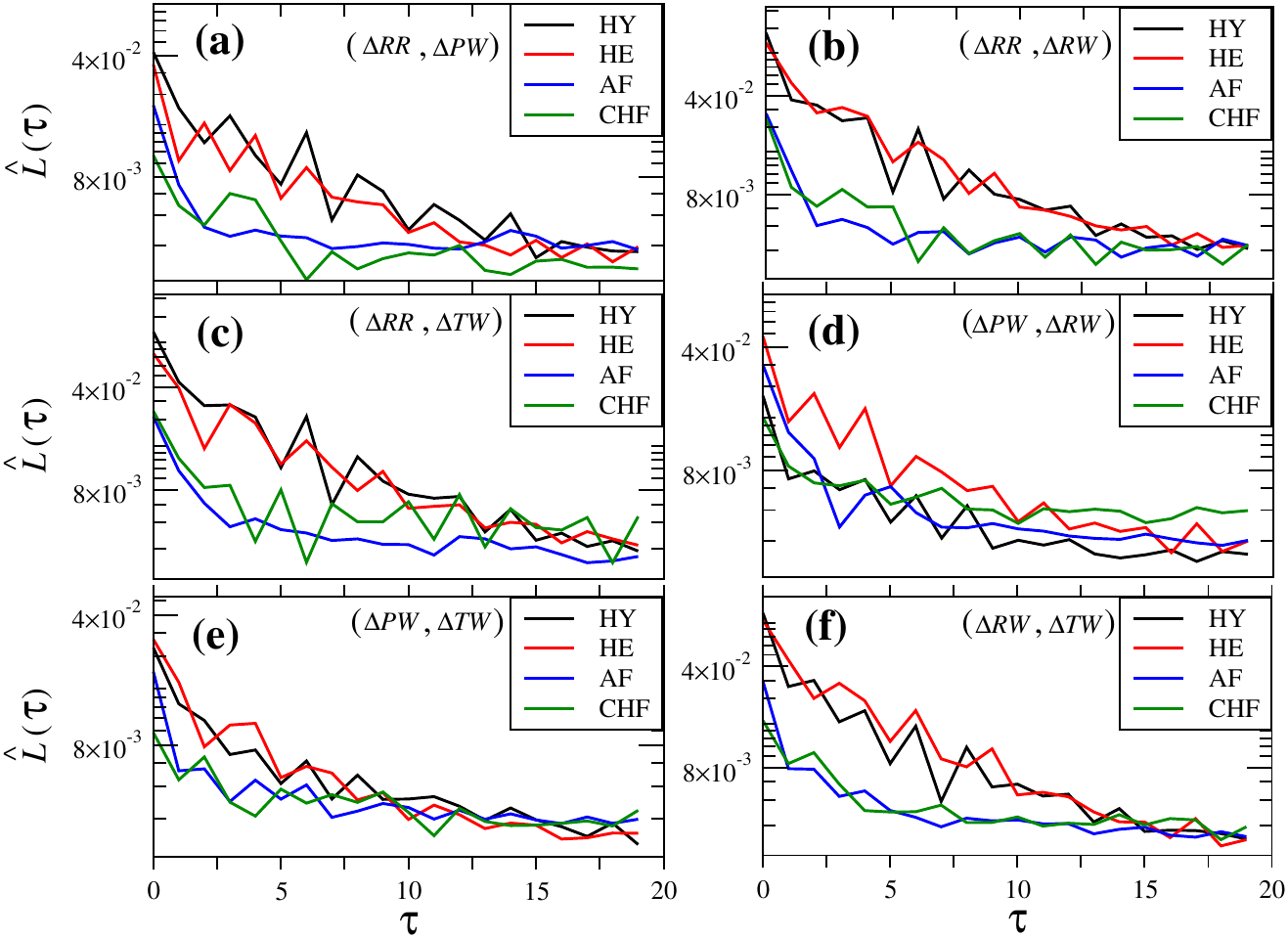}}
\end{center}
     \caption{
       Lag-irreversibility functions for joint variability signals.
 We show the estimated LI function for joint signal, i.e., we consider two 
signals obtained from an ECG to obtain  the LI function.  There are only six possible pair joint signals:  (a)  ($\Delta RR$, $\Delta PW$), (b) ($\Delta RR$, $\Delta RW$), (c) ($\Delta RR$, $\Delta TW$),(d) ($\Delta PW$, $\Delta RW$), (e) ($\Delta PW$, $\Delta TW$) and (f) ($\Delta RW$, $\Delta TW$). For each case we estimate the LI function using the same encoding scheme
as the one used in the single signal analysis, but jointly to take into account
two signals. This encoding procedure results in a time series with a states space of nine symbols 
(see Section~\ref{sec:methodology}).
            }
\label{fig:fig02-R}
\end{figure}

In Fig.~\ref{fig:fig02-R} we can see the resulting LI functions estimated from the six cases of pair variability signals: (a)  ($\Delta RR$, $\Delta PW$), (b) ($\Delta RR$, $\Delta RW$), (c) ($\Delta RR$, $\Delta TW$),(d) ($\Delta PW$, $\Delta RW$), (e) ($\Delta PW$, $\Delta TW$) and (f) ($\Delta RW$, $\Delta TW$). 
In order to estimate the LI function, we first performed the symbolic encoding  scheme described in Section~\ref{symbolicEnconding}.  
For every pair of variability signals we obtained a symbolic time series made up of nine symbols, which we used to estimate the LI function using the method  described in Section~\ref{estimationprocedure}. 
In every panel of  Fig.~\ref{fig:fig02-R} we show the average LI function for every case of joint encoding. We can qualitatively observe in that figure that almost all cases allow us to distinguish between healthy individuals (young and elderly) from those individuals with an adverse health condition. A clear exception is observed in  Fig.~\ref{fig:fig02-R} (d) where the analysis of the pair ($\Delta PW$, $\Delta RW$) fail to achieve a satisfactory discrimination between health conditions. 

In order to quantitatively evaluate how accurate is this method for discriminating signals according to the health condition, we carried out an analysis through the so-called ``Receiver operating characteristics graphs'', a technique that is referred to as \textit{ROC analysis}~\cite{fawcett2006introduction}.

%
%

\subsection{ROC analysis}

\begin{figure}[t]
\begin{center}
\scalebox{0.55}{\includegraphics{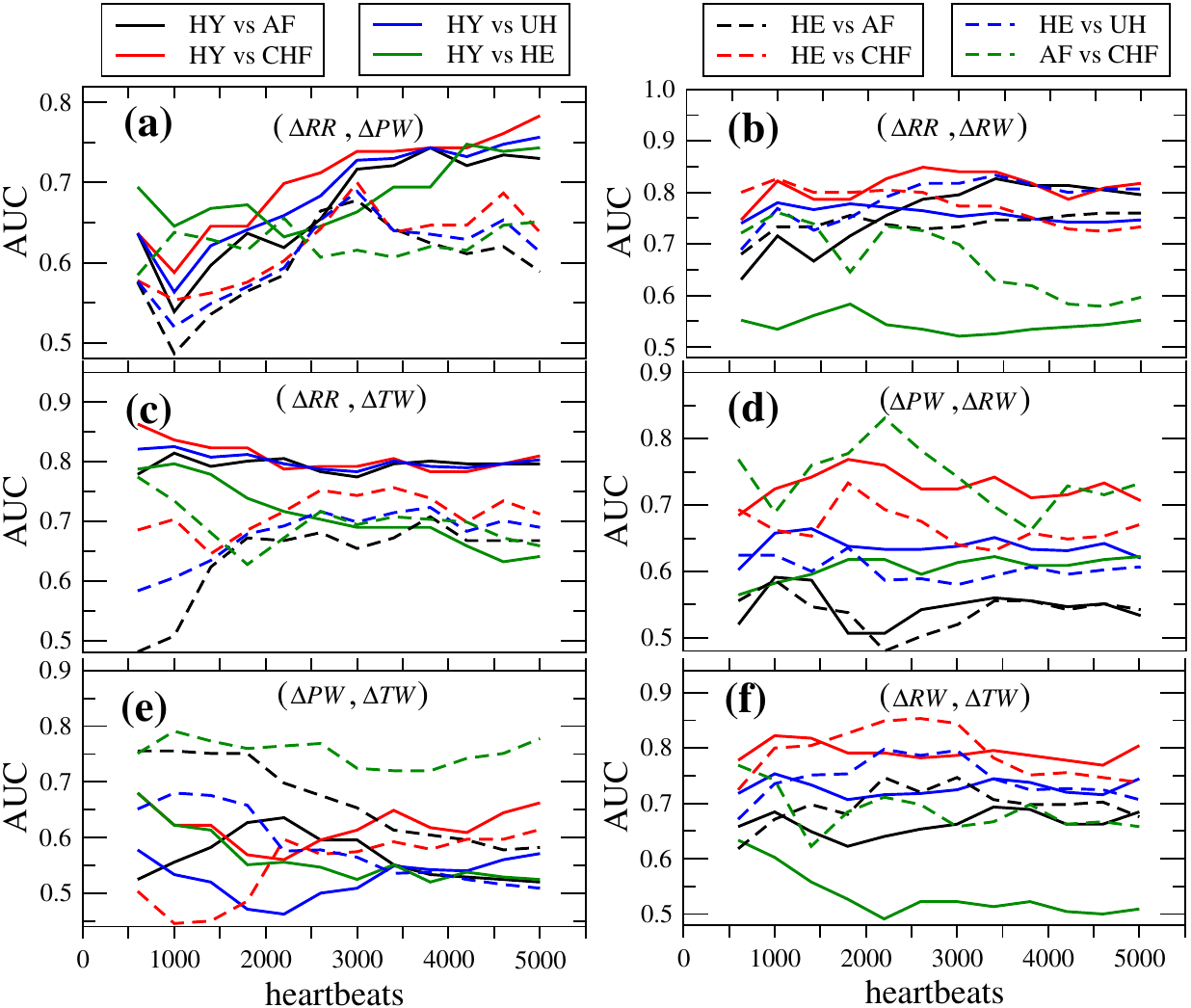}}
\end{center}
     \caption{       
       ROC analysis for the joint variability signal analysis.
            }
\label{AUC_vs_time}
\end{figure}

In this section we perform the ROC analysis, which is a binary method that allows to evaluate classifiers by means of a graph whose $x$- and $y$-axes represent the false-positive rate (denoted by g) and the true-positive rate (denoted by h), respectively~\cite{fawcett2006introduction}. To obtain the ROC curve, one chooses a threshold value of $\widehat{L}(\tau)$, above which the patient is predicted as a negative case (healthy, for instance) and below that threshold is considered to be as a positive case (unhealthy); that is, each threshold value would provide a proportion of true positive and true negative cases and their respective error proportions (false positive or false negative). Here we use this methodology to discriminate between eight pairs of patient groups, namely, H vs AF, H vs CHF, HY vs UH, HY vs HE,  HE vs AF, HE vs CHF, HE vs UH and AF vs CHF; the group UH (unhealthy group), is made up of patients exhibiting any of the adverse health conditions, AF or CHF. With this analysis, we can evaluate the performance of the method by means of {\em area under the ROC curve} (AUC)~\cite{fawcett2006introduction,mason2002areas}. The AUC quantifies the area under the curve derived by plotting $h$ vs $g$, and it takes values in the range $[1/2,1]$; the closer the values of AUC are to 1, the greater the efficiency of the method to discriminate between the two groups.  On the other hand,  for practical reasons, we are interested in having the highest value of AUC, using the shortest signal time as possible. Consequently, in Figure \ref{AUC_vs_time}, we show the AUC values as a function of time (heartbeats), for the eight pairs of patient groups and the six coding cases.

%
%
%

In Figure \ref{AUC_vs_time}, we can see that with cases ($\Delta RR$, $\Delta PW$), ($\Delta RR$, $\Delta RW$) and ($\Delta RR$, $\Delta TW$), which involve heart rate variability ($\Delta RR$), it is possible to distinguish more clearly between the two groups of healthy patients (HY and HE) and those with medical conditions. Specifically, the two pairs of groups that show the greatest area under the curve (AUC$ > 0.80$) are HY vs AF, HY vs CHF and HY vs UH, i.e.,  the methodology makes it possible to better distinguish between the group of healthy young patients from those with any of the medical conditions. The two pairs of groups that yield the lowest AUC value ($\approx 5.5$) are HY vs HE and AF vs CHF. In cases ($\Delta RR$, $\Delta RW$) and ($\Delta RR$, $\Delta TW$), the maximum AUC values ($> 0.8$) can be obtained using approximately $3\,000$ heartbeats. Regarding cases ($\Delta PW$, $\Delta RW$), ($\Delta PW$, $\Delta TW$) and ($\Delta RW$, $\Delta TW$), they yield AUC values similar to the previous cases, but to distinguish between the unhealthy groups. Explicitly, the case ($\Delta PW$, $\Delta TW$) allows reaching AUC values close to $0.8$, to distinguish between the two groups with medical conditions.

The results described above are shown quantitatively in Tables~\ref{AUC-joint-DeltaRR} and~\ref{AUC-joint-DeltaOthers}, which show the AUC values obtained when using the full signal ($5\,000$ heartbeats). We can see that values close to $0.80$ or greater are reached with pairs HY vs CHF, HY vs AF and HY vs UH, when joint coding includes heart rate variability (($\Delta RR$, $\Delta PW$), ($\Delta RR$, $\Delta RW$), ($\Delta RR$, $\Delta TW$)). In other words, we can say that, within the accuracy of our statistical analysis, this methodology is good enough for distinguishing between these groups of patients (up to $80\%$ of the analyzed cases). An interesting case not yet reported in the literature is the group of HE patients. We see that our methodology makes it possible to discriminate this group from  healthy young or individuals with AF or CHF up to $70\%$ of the analyzed cases (AUC$>0.70$). As mentioned above, these maximum AUC values can be achieved using approximately $3\,000$ heartbeats, which, in real time, is approximately 30 minutes of electrocardiographic signal.

\begin{table}[h]
\begin{center}
\begin{tabular}{c|c|c|c|}

\cline{2-4}
    &\multicolumn{3}{ |c| }{AUC} \\
\cline{2-4}
 &{\scriptsize $\Delta RR,\Delta PW$} & {\scriptsize $\Delta RR,\Delta RW$} & {\scriptsize$\Delta RR,\Delta TW$} \\
 \hline
\multicolumn{1}{|c|}{HY vs AF   } & 0.7300 & 0.8266 & 0.7966 \\
\hline
\multicolumn{1}{|c|}{HY vs CHF  } & 0.7833 & 0.8488 & 0.8300  \\
\hline
\multicolumn{1}{|c|}{HY vs UH   }& 0.7566 & 0.7466 & 0.8033 \\
\hline
\multicolumn{1}{|c|}{HY vs HE   }& 0.7433 & 0.5722 & 0.6611\\
\hline
\multicolumn{1}{|c|}{HE vs AF   }& 0.6200 & 0.7600 & 0.6677\\
\hline
\multicolumn{1}{|c|}{HE vs CHF   }& 0.6866 & 0.7733 & 0.7344\\
\hline
\multicolumn{1}{|c|}{HE vs UH   }& 0.6133 & 0.8066 & 0.7011\\
\hline
\multicolumn{1}{|c|}{AF vs CHF  }& 0.6511 & 0.6188 & 0.6988\\
\hline
\end{tabular}
\caption{ Area under the ROC curve for joint variability signals with $\Delta$RR, using $\tau=1$.}
\label{AUC-joint-DeltaRR}
\end{center}
\end{table}

\begin{table}[h]
\begin{center}
\begin{tabular}{c|c|c|c|}

\cline{2-4}
    &\multicolumn{3}{ |c| }{AUC} \\
\cline{2-4}
 & {\scriptsize$\Delta PW,\Delta RW$} & {\scriptsize $\Delta PW,\Delta TW$} & {\scriptsize $\Delta RW,\Delta TW$ }\\ 
 \hline
\multicolumn{1}{|c|}{HY vs AF   } & 0.5333 & 0.5200  & 0.6844 \\
\hline
\multicolumn{1}{|c|}{HY vs CHF } & 0.7333  & 0.6622 & 0.8044 \\
\hline
\multicolumn{1}{|c|}{HY vs UH } & 0.6422 & 0.5711 & 0.7444 \\
\hline
\multicolumn{1}{|c|}{HY vs HE   } & 0.6222 & 0.5244 & 0.5088 \\
\hline
\multicolumn{1}{|c|}{HE vs AF  } & 0.5555 & 0.5822 &  0.7022 \\
\hline
\multicolumn{1}{|c|}{HE vs CHF } & 0.6711 & 0.6144 & 0.7555 \\
\hline
\multicolumn{1}{|c|}{HE vs UH   } & 0.6066 & 0.5377 & 0.7244 \\
\hline
\multicolumn{1}{|c|}{AF vs CHF} & 0.7333 & 0.7777 & 0.6666 \\
\hline
\end{tabular}
\caption{ Area under the ROC curve for joint variability signals, using $\tau=1$.}
\label{AUC-joint-DeltaOthers}
\end{center}
\end{table}

\section{Conclusions}\label{sec:concluding}
The main purpose of this this work is twofold. Firstly, we have introduced a method for assessing time-irreversibility from time series. This method is based on the concept of pairwise reversibility, which is essentially a symmetry property of the joint distribution of a lagged pair of state variables. Based on this property we define an ``index'' to quantify irreversibility, which we called lag-irreversibility (LI) function, which is the cornerstone of our analysis. We showed through a simple example of three-states Markov chain, how  the LI function measures the degree of irreversibility of the process. In particular we argued that whenever the process is reversible the LI function, $L(\tau)$, is zero for all $\tau \in \mathbb{N}$, and that it must be positive if the process is irreversible. 

Secondly, we proceeded to apply our method. That is, we evaluated the time-irre\-ver\-sibility of time series obtained from electrocardiograms of individuals grouped into four different categories depending on their health condition: $i$) healthy young, $ii$) healthy elderly, $iii$) atrial fibrillation  and $iv$) congestive heart failure. Every ECG of each health condition was processed to obtain four variability signals:  the heart rate, the P-wave,  the R-wave,  and  the T-wave amplitude variabilities. Thus the ECG of each individual gave us four time series, which were used to perform the irreversibility analysis by means of the LI function. 
We showed that the lag-irreversibility function of every single variability signal was not efficient enough to discriminate among the different groups. Then we turn to the joint analysis of the variability signals, where it relies the main advantage of our method. For this purpose, we consider pairs of variability signals in order to evaluate the irreversibility as a joint process. We used a symbolization scheme of three states for every signal, which resulted in a symbolic time series of nine symbols (considering all the possible combinations). This time series was used to estimate the LI function for every group. We thus showed that the lag irreversibility function was able to distinguish between healthy individuals (groups $i$ and $ii$) from individuals with an adverse health condition (groups $iii$ and $iv$). The latter was made through a ROC ana\-lysis, which allows to determine the efficiency and sensibility of our test by a statistical analysis of false positives and negatives.

\section*{Acknowledgements}  
NM and CM thank CInC-UAEM for the warm hospitality during a short visit where part of the work was done. CM was financially supported by CONAHCYT grant Ciencia B\'{a}sica (Ciencia de Frontera) No. A1-S-15528. RSG thanks IPICyT for its  hospitality during academic visits and thanks CONAHCYT by partial financial support through grant number CF-2019-1327701.

\bibliographystyle{elsarticle-num} 
\bibliography{bibliography}

\end{document}